\documentclass{article}
\pdfoutput=1
\usepackage[hidelinks]{hyperref}
\usepackage[utf8]{inputenc}
\usepackage{fullpage}
\usepackage{natbib}
\usepackage{graphicx}
\usepackage{amsmath,amssymb}
\DeclareMathOperator{\E}{\mathbb{E}}
\DeclareMathOperator{\err}{err}
\DeclareMathOperator{\ed}{ed}

\date{}
\begin{document}

\title{Fast Characterization of Segmental Duplications in Genome Assemblies}

\author{
  Ibrahim Numanagić\,$^{1,2}$, 
  Alim S. Gökkaya\,$^{3}$, 
  Lillian Zhang\,$^{1}$, 
  Bonnie Berger\,$^{1,2}$, \\
  Can Alkan\,$^{3,*}$ and 
  Faraz Hach\,$^{4,5,}$
  \footnote{to whom correspondence should be addressed:
    \href{mailto:faraz.hach@ubc.ca}{faraz.hach@ubc.ca} and
    \href{mailto:calkan@cs.bilkent.edu.tr}{calkan@cs.bilkent.edu.tr}.
  }\\          
  \footnotesize
  $^{1}$ Computer Science and AI Lab, Massachusetts Institute of Technology, Cambridge 02139, MA, USA\\
  \footnotesize
  $^{2}$ Department of Mathematics, Massachusetts Institute of Technology, Cambridge 02139, MA, USA\\
  \footnotesize $^{3}$ Department of Computer Engineering, Bilkent University, 06800, Ankara, Turkey\\
  \footnotesize $^{4}$ Vancouver Prostate Centre, V6H 3Z9, Vancouver, Canada\\
  \footnotesize $^{5}$ Department of Urologic Sciences, University of British Columbia, V5Z 1M9, Vancouver, Canada\\
}

\maketitle

\begin{abstract}
{\bf Motivation:}  
Segmental duplications (SDs), or low-copy repeats (LCR), are segments of DNA greater than 1 Kbp with high sequence identity that are copied to other regions of the genome. SDs are among the most important sources of evolution, a common cause of genomic structural variation, and several are associated with diseases of genomic origin including schizophrenia and autism. Despite their functional importance, SDs present one of the major hurdles for \textit{de novo} genome assembly due to the ambiguity they cause in building and traversing both state-of-the-art  overlap-layout-consensus and de Bruijn graphs. This causes SD regions to be misassembled, collapsed into a unique representation, or completely missing from assembled reference genomes for various organisms.
In turn, this missing or incorrect information limits our ability to fully understand the evolution and the architecture of the genomes. Despite the essential need to accurately characterize SDs in assemblies, there is only one tool that has been developed for this purpose, called Whole Genome Assembly Comparison (WGAC); its primary goal is SD detection. WGAC is comprised of several steps that employ different tools and custom scripts, which makes this strategy difficult and time consuming to use. Thus there is still a need for algorithms to characterize within-assembly SDs quickly, accurately, and in a user friendly manner.

{\bf Results:}  
Here we introduce SEgmental Duplication Evaluation Framework (SEDEF) to rapidly detect SDs through sophisticated filtering strategies based on Jaccard similarity and local chaining. We show that SEDEF accurately detects SDs while maintaining substantial speed up over WGAC that translates into practical run times of minutes instead of weeks. Notably, our algorithm captures up to $25\%$ ``pairwise error'' between segments, whereas previous studies focused on only $10\%$, allowing us to more deeply track the evolutionary history of the genome. 

{\bf Availability:} SEDEF is available at \href{https://github.com/vpc-ccg/sedef}{https://github.com/vpc-ccg/sedef} 
\end{abstract}

\section{Introduction}
Segmental duplications (SDs) are defined as genomic segments of size greater than 1 Kbp that are repeated within the genome with at least 90\% sequence identity~\citep{Bailey2001} in either tandem or interspersed organization. Almost all genomes harbor large SDs; for example the build 37 release of the human reference genome (GRCh37) contains a total of 159 Mbp gene-rich duplicated sequence, which corresponds to approximately 5.5\% of its entire length\footnote{\href{http://humanparalogy.gs.washington.edu/build37/build37.htm}{http://humanparalogy.gs.washington.edu/build37/build37.htm}}. 
It is known that SDs played a major role in evolution~\citep{Marques-Bonet2009,Prado-Martinez2013,Sudmant2013},
and are one of the most important factors that contribute to human disease either directly~\citep{Gonzalez2005,Yang2007,Hollox2008}, or through leading to other forms of structural variation~\citep{Alkan2011,Mills2011}. Furthermore, human populations show SD diversity that may be used as markers for population genetics studies~\citep{Alkan2009,Sudmant2010}.

Despite their functional importance, SDs are poorly characterized due to the difficulties they impose on constructing accurate genome assemblies~\citep{Alkan2011b,Chaisson2015,Steinberg2017}, as well as the ambiguities in read mapping~\citep{Treangen2012,Firtina2016}. Building {\em inaccurate\/} assemblies due to SDs is an important problem, since it may lead to potentially incorrect conclusions about the evolution of the species of interest as in the case of the giant panda assembly~\citep{Li2010c}. In the giant panda genome analysis, the authors concluded that the panda genome included substantially less repeats and duplications compared to other mammalian genomes, however, it is shown that this result is likely incorrect due to mis-assembly of these regions.
Additionally, duplications give rise to gaps in the assembly, which likely contain genes and other functionally active regions~\citep{Bailey2001,Bailey2002,Alkan2011b}.

Accurate assembly of duplicated regions remains a difficult and unsolved problem, which may be ameliorated through the use of ultra-long reads generated by the Oxford Nanopore platform~\citep{Jain2018} 
or Linked-Read sequencing~(e.g., 10x Genomics \citep{Mostovoy2016,Yeo2018}) if the duplicated segment is shorter than the read or linked-read length, respectively. 
However, characterization of the SD content in existing assemblies is still important for two reasons: (i) to evaluate the ``completeness'' of these genome assemblies, and (ii) understand genome evolution for comparative genomics studies.

SD content in assemblies can be assessed using two strategies, and the overlap between the results of the two methods determines the completeness of the assembly in terms of duplications. The first method, called Whole Genome Assembly Comparison (WGAC), relies on the alignment of the entire genome to itself to identify repeating segments~\citep{Bailey2001} within the assembly, except for common repeats which are filtered out. The second strategy is called Whole-genome Shotgun Sequence Detection (WSSD) which relies on the read depth sequence signature~\citep{Bailey2002}. Briefly, the WSSD method aligns the original reads back to the assembled genome, and looks for regions of read depth significantly higher than the average, which signals a putative duplication~\citep{Bailey2002}. 
Regions that are marked as SDs by WSSD, but not by WGAC, are then classified as likely collapsed duplications~\citep{Alkan2011b}.


Although the ``optimal'' alignment of the entire genome to itself can be theoretically computed via standard dynamic programming  (e.g. Smith-Waterman algorithm), such an approach remains impractical due to quadratic time and memory complexity, and is likely to remain so \citep{Backurs2015}. Furthermore, high edit distance between the SD paralogs disqualifies the use of most of the available edit distance approximations with theoretical guarantees \citep{Andoni2010,Hanada2011}, as well as the majority of the sequence search tools that operate under the assumption that the similarity between regions is high. 
Standard whole-genome or long-read aligners (e.g. MUMmer~\citep{marccais2018mummer4} or Minimap2~\citep{Li2018}) are not able to efficiently capture SDs with low similarity rates (i.e. lower than 80\%), and also confuse SDs with other repeated elements in the genome (e.g. stretches of short tandem repeats).
For these reasons, the WGAC method is composed of a number of heuristics that include several tools and scripts~\citep{Bailey2001}. First, the common repeats are removed from the assembly in a step called \textit{fuguization}. Remaining regions are partitioned into 400 Kb chunks (due to memory limitations when WGAC was developed), all pairwise alignments are computed using BLAST~\citep{Altschul1990}, and significant alignments are kept as putative duplications.  A modified version of BLASTZ~\citep{Schwartz2003} is also used to find within-chunk (i.e. 400 Kb segments) duplications.
Next, common repeats are inserted back, spurious alignments at the end of the sequences are trimmed, and final global alignments are calculated.\footnote{Another method, SDquest~\citep{Pu2018}, was published while this manuscript was under review.}

The original WGAC implementation as outlined above is difficult to run, and it is time consuming as it relies on several general purpose tools (such as BLAST) and custom Perl scripts. In its current form, the only way to accelerate the WGAC analysis is using a compute cluster to parallelize BLAST alignments. Interestingly, another problem is the modified version of BLASTZ for self alignments: the source code is not available, and only a binary compiled for the Sun Solaris operating system has been released\footnote{\href{http://humanparalogy.gs.washington.edu/code/WGAC\_HOWTO.pdf}{http://humanparalogy.gs.washington.edu/code/WGAC\_HOWTO.pdf}}, rendering the tool unusable for most other researchers\footnote{We note that this self-alignment step might be replaced with another tool such as LASTZ~\citep{Harris2007}, however, the parameter settings for the current release is not yet optimized for alternative aligners.}.

Here we introduce a new algorithm to characterize SDs in genome assemblies. While we follow a strategy akin to WGAC in aligning a whole genome to itself, we do so in a more efficient way by introducing sophisticated optimizations to both the putative SD detection and global alignment steps.
We leverage our knowledge from the biology of human and other genomes that different mutation events contribute unequally to the total value of the error rate (quantification of differences) between segments, in order to better optimize our SD detection algorithms.
A key conceptual advance of our work 
that helps model such events in the genome is to separately consider germline mutation rates (denoted as \emph{small mutations}), and larger-scale \textit{de novo} structural variation (SV) rates. This formulation enables us to speed up SD detection and better capture evolutionary events.
We implement our algorithms in C++ and provide a single package called SEDEF (SEgmental Duplication Evaluation Framework). In contrast to WGAC, which requires several weeks to complete even on a compute cluster, SEDEF can characterize SDs in the human genome in 10 CPU hours.
We believe SEDEF will be a powerful tool to characterize SDs for both genome assembly evaluation and comparative genomics studies.   


\section{Preliminaries}
Segmental duplications (SDs) are generated by large-scale copy events that have occurred during the 
evolution of the genome. After such a copy event, both sites involved in SD may have undergone a number
of changes during the evolutionary history of the genome. 
Formally, consider a genomic sequence $G=g_1g_2g_3\dots g_{|G|}$ of length $|G|$, where $g_i \in \{A, C, G, T\}$ for any $i$. 
Let $G_{i:i+n}=g_i,\dots,g_{i+n-1}$ 
be a substring in $G$ of length $n$ that starts at position $i$. 
Furthermore, let $X_i$ be the set of all  $k$-mers in the substring $G_{i:i+n}$. We assume that $k$ is predefined and fixed. 

The Levenshtein~\citep{Levenshtein1966} \emph{edit distance} is defined as: $\ed(G_{i:i+n},$ $G_{j:j+m})$ of two substrings $G_{i:i+n}$ and $G_{j:j+m}$ (further simplified as $G_i$ and $G_j$) to be a minimal number of edit operations (i.e. single nucleotide substitutions, insertions, and deletions) that are needed to convert the string $G_{i}$ into $G_{j}$. The length of the alignment between $G_i$ and $G_j$ is denoted as $l$, and clearly $l\geq \max(m,n)$.
Let us define the notion of \emph{edit error} (further referred to as just \emph{error}) between two strings $G_i$ and $G_j$ as 
$\err(G_i, G_j)=\ed(G_i, G_j)/l$---the edit distance normalized over alignment length. 
Intuitively, this is the average number of the edits needed to turn $G_i$ into $G_j$.
Clearly, two strings are identical if $\err(G_i, G_j)=0$. 
We consider $G_i$ and $G_j$ as a \emph{segmental duplication} of length $l$ with error $\delta$ if the following \emph{SD conditions} are met:
\begin{itemize}
\item $l \geq 1,000$ where $l$ is the length of alignment between $G_i$ and $G_j$,
\item $\err(G_i, G_j)\leq\delta$.
\end{itemize}
We also assume that the overlap between the $G_i$ and $G_j$ in the genome is at most $\delta \cdot n$.

\subsection{Edit distance model}
\label{sec:sdmodel}

Each segmental duplication is generated by a past structural variation (SV) event that copied a substring of length $n$ in $G$ from locus $i$ to locus $j$. This copy was initially perfect, meaning that corresponding strings $G_i$ and $G_j$ (initially both of length $n$) were identical. However, various changes during the evolutionary history of the genome--- such as point mutations, small indels, and  other structural variants--- have altered both the original and duplicate strings independently. Thus it is necessary to take such changes into consideration when identifying the potential SDs. While previous SD studies focused only on SDs with pairwise error at most $10\%$, here we aim to focus on SDs whose error rate can go up to $25\%$ (in another words, $\delta\leq 1/4$). Higher $\delta$ allows us to track the evolutionary history of the human genome to earlier periods. However, it also significantly renders the SD detection problem more difficult since the majority of the known filtering techniques that operate on the edit distance metric space assume much lower values of $\delta$ \citep{Andoni2010}. We address this challenge by leveraging our knowledge from the biology of human and other genomes that different mutation events contribute unequally to the total value of $\delta$ in order to better optimize our SD detection algorithms.

\begin{figure*}
\centering
\includegraphics[scale=0.24]{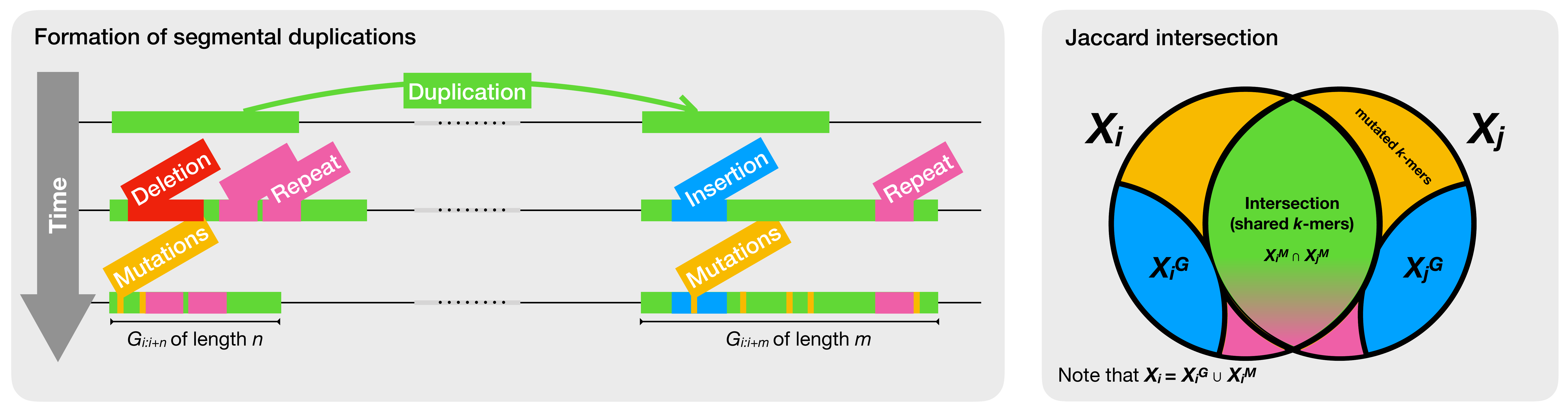}
\caption{{\bf (Left)\/} Simplified representation of a segmental duplication lifetime. Initially, a large-scale duplication forms an SD, at which point both the original region and the copy are identical. Then, both the original region and copy undergo various independent changes, such as large-scale deletions (in red), insertions (in blue), and small repeat insertions (in fuchsia). Finally, various germline mutations (in yellow) affect both regions. The resulting SD as seen today, defined as the pair $(G_{i:i+n},G_{j:j+m})$, is shown in the third row. {\bf (Right)\/} Shows the idealized Jaccard similarity between the $k$-mer sets $X_i$ and $X_j$ corresponding to the $G_i$ and $G_j$, respectively. Note that some repeats also increase the proportion of shared $k$-mers. Colors denote same as on Left.}
\label{fig:model}
\end{figure*}

A key conceptual advance of our work that helps model such events in the human genome is to separately consider germline mutation rates (denoted as \emph{small mutations}), and \textit{de novo} SV rates. It is estimated that the 
substitution rate in the human genome is roughly $0.5 \times 10^{-9}$ per basepair per year \citep{Scally2016}, and we may assume a similar rate for other mammalian genomes. The evolutionary split of the human and chimpanzee species  
is estimated to have occurred approximately 7 million years ago \citep{Hedges2009}. Thus, we expect that the probability of a basepair being mutated since the split is roughly $3.5 \times 10^{-3}$. If we also account for small indels (with an  even smaller mutation rate than those of substitutions~\citep{Montgomery2013}), the edit error between any two paralogs of an SD that have occurred after the evolutionary split is not larger than 0.1\%. Even if we consider SDs that occurred much earlier in history (e.g. after the lowest common ancestor of human and mouse roughly 90 million years ago), the total edit error will not be larger than 10\%. 
However, the edit error between two paralogs of an SD can be much larger due to large structural variations. One such example is insertion of transposons within an SD. These events can be visualized as large gaps within an edit distance string of two segmental duplications which contribute a large share towards the total edit error (Figure~\ref{fig:model})\footnote{We note that inversion and translocation events contribute more to sequence divergence due to incorrect alignments, but such events are rare.}. 
Thus we assume that small mutations contribute at most $\delta_M \leq 0.15$ towards the total edit error $\delta$ (both paralogs can be mutated up to 7.5\%); this default setting is higher than the estimated rates (above) for human and mouse genomes in order to be able to handle older species as well. Analogously, large-scale events (subsequently referred to as \emph{gaps}) contribute the remaining $\delta_G = \delta - \delta_M$ of the edit error. We assume that the probability of a large gap occurring at any basepair in the genome is not larger than $0.005$ (as estimated by analysis of existing human SDs; similar value can be derived for other species). 
Note that the gap penalty is typically calculated via affine gap model, where gap openings are heavily penalized while the gap extensions are either ignored or assigned a very low penalty. Many human SDs have $\delta_G \gg 0.15$ if calculated by standard Levenshtein distance metric. SEDEF uses standard (Levenshtein) gap distance metric while locating seed SDs (meaning all seed SDs have $\delta_G\leq 0.15$); however, this restriction is lifted in a later step where we switch to the affine gap penalty.

Furthermore, we assume that the mutations within SD paralogs follow a Poisson error model \citep{Jain2017a, Fan2015}, and that those mutations occur independently of each other. It follows that any $k$-mer in $X_j$ (the set of $k$-mers of $G_j$) has accumulated on average $k\cdot\delta_M$ mutations compared to the originating $k$-mer in $X_i$ provided that such a $k$-mer was part of the original copy. By setting a Poisson parameter $\lambda=k\cdot \delta_M $, we obtain the probability of an event in which a $k$-mer is preserved in both paralogs of an SD (i.e. that it is error-free): $\Pr(\text{\# mutations} = 0\mid\lambda)=e^{-k \delta_M}$. 

We call this error model \emph{SD error model}, and assume that any SD of interest satisfies the error constraints mentioned above.

\subsection{Jaccard similarity}
\label{sec:jaccard}

Suppose that we ask whether two substrings $G_i$ and $G_j$ are similar to each other, where the length of both strings is $n$. One way to measure the similarity of those substrings is to analyze their respective $k$-mer sets $X_i$ and $X_j$ and to count the number of shared $k$-mers between them. This metric is known as \emph{Jaccard similarity} of sets $X_i$ and $X_j$, and is formally defined as 
$$J(X_i,X_j)=\frac{|X_i\cap X_j|}{|X_i\cup X_j|}.$$ 
Clearly, higher similarity of strings $G_i$ and $G_j$ implies a larger value of $J(X_i,X_j)$. 

The calculation of Jaccard similarity between two sets can be approximated via the MinHash technique developed by Broder \citep{Broder1997}, who proved that given a universe of all $k$-mers $U$ and a random permutation $h$ on $U$ (typically a  hash function with no collisions), it follows that
$J(X_i,X_j) = $ $\Pr \left[ h_{min} (X_i) = h_{min} (X_j)\right]$ where $h_{min}(T)$ is the minimal member of $T$ with respect to $h$.
Furthermore,  $J(X_i,Y_j)$ can be estimated and efficiently computed by calculating
$$\frac{\left|S(X_i \cup X_j) \cap S(X_i) \cap S(X_j)\right|}{\left|S(X_i \cup X_j)\right|}$$
where $S(X_i)$ is the \emph{sketch of $X_i$} and stands for a subset of $s$ elements from $X_i$ whose hash values are minimal with respect to the hash function $h$ (such elements are called \emph{minimizers} of set $X_i$). This estimate is unbiased as long as $h$ is random, and its accuracy depends on the sketch size $s$. Since in practice $s$ is much smaller than $|X_i|$, calculating a MinHash estimate is substantially faster compared to the calculation of $J(X_i,X_j)$. 

The performance of the MinHash technique can be further improved in the context of large strings, as shown by Jain \emph{et al}~\citep{Jain2017a}.
Instead of computing $S(X_i)$, it is possible to compute $S(W({X}_i))$, where $W({X}_i)$ is a \emph{winnowing fingerprint} of the corresponding string $G_i$. $W({X}_i)$ is calculated by sliding a window of size $w$ through $G_i$ and by taking in each window a $k$-mer of minimal hash value (in case of a tie, the rightmost $k$-mer is selected). The expected size of $W(A)$ for a random sequence $A$ is $2|A|/(w+1)$ \citep{Schleimer2003}. The main benefit of winnowing, aside from speeding up the construction of sketch $S(A)$, is the fact that winnow $W(A)$ can be computed efficiently in linear time and $O(w)$ space in a streaming fashion with appropriate data structures \citep{SmithWW}.

Moreover, it has been empirically shown that $J(X_i,X_j)$ can be efficiently estimated by calculating a \emph{winnowed MinHash} score of sets $X_i$ and $X_j$ \citep{Jain2017a}:
$$\frac{|S\left[W({X}_i) \cup W({X}_j)\right] \cap S\left[W({X}_i)\right] \cap S\left[W({X}_j)\right]|}{|S\left[W({X}_i) \cup W({X}_j)\right]|}.$$
Furthermore, given the minimal desired value $\tau$ of Jaccard similarity between the two sets,
it follows that $|W({X}_i) \cap W({X}_j)| \geq s \cdot \tau$ \citep{Jain2017a}. This estimate can be used to efficiently filter out any sets $X_i$ and $X_j$ whose Jaccard similarity is below a given threshold with high confidence.

\section{Methods}

The segmental duplication detection problem can be formulated as follows:
find all pairs of loci $(i, j)$ inside a genome $G$ with $l \geq 1,000$
such that (i) the edit error (i.e. divergence) between $G_i$ and $G_j$ is at most $\delta$; and (ii) the corresponding alignment between $G_i$ and $G_j$ is of size at least $l$ and is not contained within a larger alignment satisfying the SD criteria. In other words, for any pair $(i,j)$ we aim to find a maximal valid region alignment of size $l$ between $G_i$ and $G_j$ that satisfies the criteria and error model of segmental duplications. 

\begin{figure*}
\centering
\includegraphics[scale=0.23]{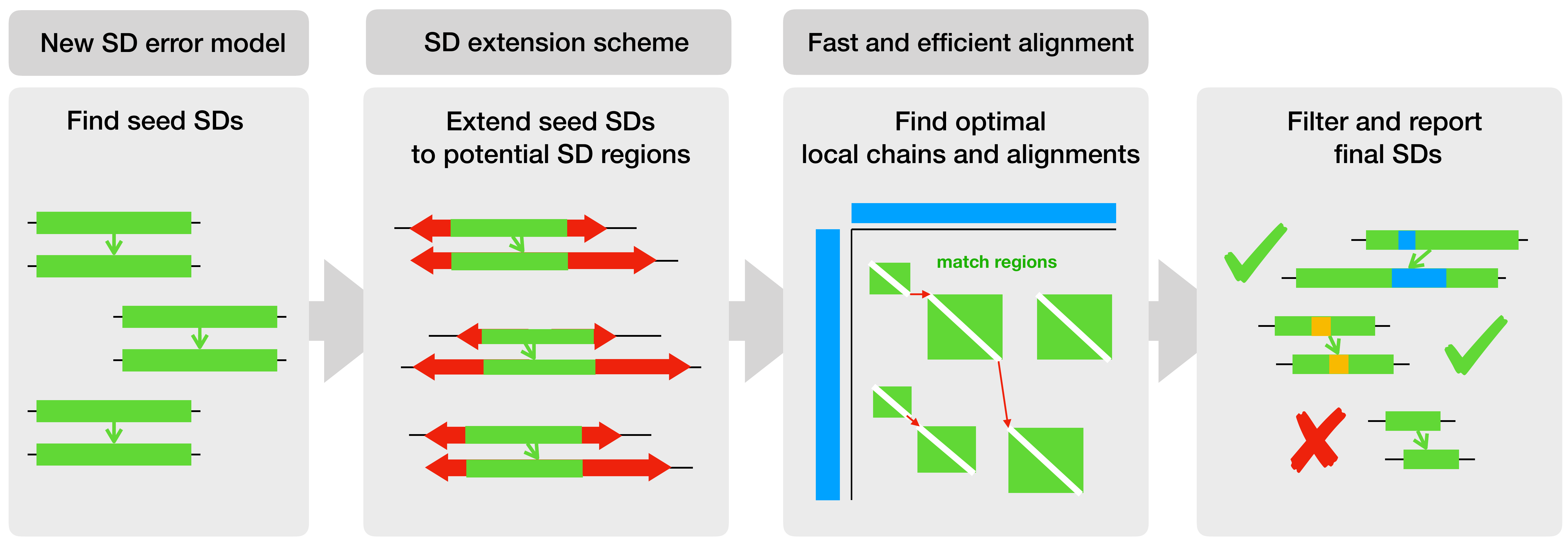}
\caption{Step-by-step depiction of the SEDEF framework. Our contribution is highlighted above the steps in the dark gray boxes.}
\label{fig:pipeline}
\end{figure*}

A na\"{i}ve method for locating SDs within any genomic sequence $G$ consists of locally aligning $G$ onto itself, followed by the analysis of all acceptable paths within a local alignment matrix. However, this strategy is impractical for large $|G|$ since the best known algorithms for optimal local alignment require $O(|G|^2)$ time and space. Another possible approach, which we take, is to solve this problem by iterating through each pair of indices $(i,j)$ within $G$ and testing whether the matching $G_i$ and $G_j$ satisfy the SD criteria through global alignment (given a fixed size $n$ of $G_i$ and $G_j$). While this method, if implemented na\"{i}vely, is still too slow for larger genomes and requires quadratic space, it can be significantly accelerated by filtering out any pair $(i, j)$ that is unlikely to form an SD. This iterative approach is the cornerstone of our SD detection framework, SEDEF, which consists of a novel seed and extend algorithm (Figure~\ref{fig:pipeline}):\\
\textbf{SD seeding:} 
Initially, we aim to find all pairs of strings $(G_i,G_j)$--- called \emph{seed SD}--- such that the length of both strings is $n \geq 1000\cdot(1 - \delta) = 750$, and such that $G_i$ and $G_j$ are believed to satisfy the SD criteria. We achieve this by iterating through the genome, and for each locus $i$ in the genome rapidly enumerating all feasible pairs $j$ for which winnowed MinHash Jaccard similarity between $G_i$ and $G_j$ goes over a pre-defined threshold $\tau$. \\
\textbf{SD extension:} 
Here we relax the condition that both $G_i$ and $G_j$ have the same size $n$, and keep expanding both seed regions $G_i$ and $G_j$ until the winnowed MinHash estimate drops below $\tau$. These enlarged seed SDs are called \emph{potential SD regions}. We terminate this  extension when we either reach the maximal allowed value of SD, or if the extension causes $G_i$ and $G_j$ to significantly overlap. \\
\textbf{SD chaining:} 
Finally, we locate all ``true'' SDs within any potential SD region and calculate their alignments by locally aligning potential SD regions via local chaining and sparse dynamic programming. Afterwards, we filter out any spurious hits and report the remaining SDs.

\subsection{Identifying seed SDs}

In order to  verify if the strings $G_i$ and $G_j$ have edit error $\leq \delta$ under the SD error model, 
 we will calculate the Jaccard similarity of their corresponding $k$-mer sets $X_i$ and $X_j$ and check if it is $\geq \tau$. 
For the sake of explanation, we will assume that $n=|G_i|=|G_j|$ (analogous reasoning holds if this is not the case).
If $c$ is $|X_i \cap X_j|$ (number of shared $k$-mers), and $t=n-k+1$ (the size of sets $X_i$ and $X_j$), we 
 can express the Jaccard similarity of those sets as $J(X_i,X_j)=c/(2t-c)$. We will also assume that no $k$-mer occurs twice in these sets; this assumption is sufficient for the calculation of lower bound below.

Simply using error $\delta$ to calculate the expected lower bound of Jaccard similarity $\tau$ is infeasible
 in practice due to the large value of $\delta$ in our setting.
However, as noted above, differences between duplicated regions are not chosen uniformly at random, and thus we can separate the error $\delta$ into the $\delta_M+\delta_G$, where $\delta_M$ is the error rate of the 
 small mutations and $\delta_G$ is the error rate of large indels, as defined in the preliminaries.
This separation of the two error rates is one of the novel contributions of our work.

If there exists a valid SD spanning $G_i$ and $G_j$, set $X_j$ can be considered as a union of two disjoint sets $X^G_j$ and $X^M_j$, where $X^M_j$ represents the $k$-mers initially copied by the SD event and might have undergone small mutations, while $X^G_j$ contains all ``new'' $k$-mers introduced by subsequent large events such as SVs and large indels
(analogous separation applies to $X_i$ as well). In this way we can separate the effects of the small mutations and large-scale events. Ideally $X^G_j$ shares no $k$-mers with $X_i$, while $X^M_j\cap X^M_i\neq\emptyset$ since we expect some shared $k$-mers to remain error-free after small mutations (see Figure~\ref{fig:model} for visualization). 
Now let us also express $t$ as $t_M+t_G$, where $t_M=|X^M_j|$ and $t_G=|X^G_j|$ (note that $|X_i|=|X_j|$ implies $|X^G_i|\approx|X^G_j|$ because the small mutations keep strings that generate $|X^M_i|$ and $|X^M_j|$ similar in size; thus we assume w.l.o.g that $|X^G_i|=|X^G_j|$).

Let $c/t_M$ be the ratio of $k$-mers that are not mutated in both $X^M_i$ and $X^M_j$ (we assume that $X^G_i$ and $X^G_j$ share no common $k$-mers, which is a valid assumption for a lower bound calculation). Its expected value, provided a Poisson error model introduced above, is $\E[c/t_M] = e^{-k \delta_M}$ \citep{Jain2017a}.  

Now we proceed to estimate the minimal required Jaccard similarity $J(X_i, X_j)$ of $X_i$ and $X_j$. Note that so far:
\begin{enumerate}
\item $|X_i \cap X_j|=|X^M_i \cap X^M_j|$; 
\item $t_G/(t_M+t_G) \leq \delta_G \Rightarrow t_G \leq t_M \cdot \delta_G/(1-\delta_G)$; and
\item $|X^G_i \cup X^G_j| \leq 2|X^G_j| = 2 t_G$ because $|X^G_i|=|X^G_j|$ (equality holds for the ideal condition where $|X^G_i \cap X^G_j|=\emptyset$).
\end{enumerate}
It follows that:
\begin{align*}
J(X_i,X_j) 
    &= \frac{|X_i\cap X_j|}{|X_i \cup X_j|} =
	       \frac{|X^M_i\cap X^M_j|}{|X^M_i\cup X^M_j| + |X^G_i\cup X^G_j|} & \text{(by 1.)} & \\
	& \geq  \frac{|X^M_i\cap X^M_j|}{|X^M_i\cup X^M_j| + 2t_G} & \text {(by 3.)}  \\
    &\geq  \frac{|X^M_i\cap X^M_j|}{|X^M_i\cup X^M_j| + 
    			\frac{2 \delta_G}{1-\delta_G} |X^M_i\cup X^M_j|} & \text{(by 2.)} 
\end{align*}
\begin{align*}
       & &=     \frac{1-\delta_G}{1+\delta_G} \frac{|X^M_i\cap X^M_j|}{|X^M_i\cup X^M_j|} = \frac{1-\delta_G}{1+\delta_G} J(X^M_i, X^M_j). &
\end{align*}
Since $J(X^M_i, X^M_j)=c/(2t_M-c)$ and the expected value of $c/t_M$ is $e^{-k \delta_M}$, it clearly follows that the minimum required expectation of Jaccard similarity $\tau$ is at least:
$$
\tau = \E[J(X_i,X_j)] \geq \frac{1-\delta_G}{1+\delta_G} \cdot \frac{1}{2e^{k\delta_M}-1}.
$$

To find the seed SDs, we follow a similar strategy as described in~\citep{Jain2017a}, where our $G_i$ and $G_j$ 
correspond to the long reads and the genomic hits. 
We start by indexing a genome $G$ and constructing an index $I_{G}$ of genome $G$ that is a sorted list of unique  pairs $(i, x)$ where $x$ is a $k$-mer in the winnow $W({G})$ and $i$ is a starting position of $x$ in $G$. We also construct a reverse index  $I^{-1}_{G}$: it provides for any input $k$-mer $x$ a list of all positions $i$ in $G$ such that $(i, x) \in I_{G}$. These two tables are computationally inexpensive to calculate and allow us to quickly calculate winnow $W({X}_i)$ of any substring $G_i$ in $G$. For any locus $i$ within $G$, we enumerate a list of all pairs $C=\{(j,x) \in I_G: x \in W(G_i)\}$. By using the winnowed MinHash lemma, we know that substring $G_j$ starting at some locus $j$ is a potential SD match for $G_i$ if $W(G_i)$ and $W(G_j)$ share at least $\tau \cdot s$ $k$-mers, where the sketch size $s$ is set to $|W(G_i)|$. Since $C$ is sorted by index, we can use this lemma to efficiently select all candidate locations $j\in[j_{a}, j_{b}]$ for which $J(G_i, G_j) \geq \tau$ by ``rolling'' a MinHash calculation as follows~\citep{Jain2017a}. We start by setting $j=j_a$, and then construct an ordered set  
\begin{align*}
  & & L= \{(y, b):\: & y \in W(X_i) \cup W(X_{j}) \\
  & & & \text{ and } b = 1\text{ if }y \in W(X_i) \cap W(X_j)\},  
\end{align*}
where each element $y$ is assigned 1 if it belongs to the intersection of $W(X_i) \cap W(X_j)$ and zero otherwise.\footnote{Such a set can be efficiently implemented with a balanced binary tree where any update operation costs only $O(\log |L|)$.} Then we keep ``rolling'' $G_j$ by increasing $j$: this corresponds to checking the similarity between $G_{i:i+n}$ and $G_{j+1:j+n+1}$, wherein we remove any minimizer from $L$ which occurred at position $j$ and add any minimizer that occurs at the position $j+n+1$. Note that any such step costs at most $O(\log s)$ operations (where $s$ is the sketch size). With the appropriate auxiliary structures, we can calculate the winnowed MinHash estimate of $W(X_i)$ and $W(X_{j+1})$ in $O(1)$ time. Once we find a $j$ for which the corresponding MinHash estimate is maximal and above $\tau$, we add the pair $(i,j)$ to the list of found SD seeds. 

\subsection{Finding potential SD regions} 

So far, we have assumed that the value of $n$ is fixed and that $n=|G_{i}|=|G_{j}|$. Now we lift this restriction and attempt to extend any seed SD as much as possible in both directions in order to ensure that we can find the boundaries of ``true'' SDs. 
This can be done by iteratively increasing the values of $n$ and $m$ by one (each step takes $O(\log s)$ time), which essentially keeps expanding the sets $W(X_i)$ and $W(X_j)$: any minimizer which occurs at loci $i+n+1$ and $j+m+1$ within $G$ is added to the ordered set $L$. Here we utilize the same structures as in the previous step (see Section 3.1), and keep extending SD region until the value of the winnowed MinHash estimate goes below $\tau$. We also terminate extension if both $n$ and $m$ become too large (we limit SEDEF to find potential SDs of at most 1 Mbp in length, as per WGAC). 
Note that the term $|S(W(X_i) \cup W(X_j))|$ keeps growing while $|S(W(X_i) \cap W(X_j))|$ stays the same if two regions stop being similar after some time, which iteratively lowers the Jaccard estimate.
We also interrupt the extension if the strings $G_{i}$ and $G_{j}$ begin to overlap. Note that we can perform this extension in the reverse fashion, by slowly decreasing the values $i$ and $j$ and applying the same techniques as described above. Finally, we report the largest $G_{i}$ and $G_{j}$ whose corresponding MinHash estimates are above $\tau$.

For each potential SD, we also apply a $q$-gram filter \citep{Jokinen1991} in order to further reduce the rate of false positives as follows. Define the $q$-gram similarity $Q(G_i, G_j)$ of strings $G_i$ and $G_j$ to be the total number of $q$-mers shared by both $G_i$ and $G_j$. We adapt the well-known $q$-gram lemma for our problem as follows: any $G_i$ and $G_j$ whose edit error is below $\delta$ and satisfies the SD error model will share at least $n (1 - \delta_G - q\delta_M ) - (np_G + 1) \cdot (q - 1)$
$q$-grams, where we assume that $n\leq m$ and where $p_G$ is the expected number of gaps per basepair in the genome. This modification allows us to losslessly reject any pair of substrings $G_i$ and $G_j$ that do not satisfy the SD error model for the given value of $p_G$. 

The aforementioned algorithm performed on the whole human genome produces more than 500 million potential SD regions due to the presence of various small repeats in the genome. In order to alleviate this problem, we only use $k$-mers that contain at least one non-repeat-masked nucleotide during the detection of seed SDs. In order to allow the case of repeats being inserted in the SD during the evolutionary process, the SD extension step uses any available $k$-mer to extend seed SDs. Finally, we pad each potential SD region with a pre-defined number of bases (which is a function of the size of the potential SD region) in order to further increase the probability of locating large SDs within the potential regions.

\subsection{Detecting final SDs}

After finding the potential SD regions, we enumerate all local alignments of size $1,000$ within those regions that satisfy the SD criteria.
In order to do this efficiently, SEDEF employs a two-tiered local chaining algorithm similar to those in \citep{Abouelhoda03,Myers95}. In the first part, we use a seed-and-extend method to construct the list of matching seed locations (of size 11 and higher), and proceed by finding the longest chains formed by those seeds via an $O(n \log n)$ sparse dynamic programming algorithm as described in \citep{Abouelhoda03,Myers95}. In this step, we restrict the maximum gap size between the seeds to $l \cdot \delta_G$ in order to cluster the seeds within a chain as ``close'' as possible. After finding these \emph{initial} chains (which might span less than $1,000$bp), we \emph{refine} them by further chaining them into the large final chains by allowing larger gaps. In order to retain compatibility with WGAC, which allows arbitrary large gaps within the SD (since it does not penalize the gap extension), we use the affine gap penalty during the construction of SD chains; however, we limit gaps to no longer than $10,000$bp in order to avoid low-quality alignments. Chaining is accompanied by the global alignments which are done with the KSW2 library, which utilizes ``single instruction, multiple data'' (SIMD) parallelization through SSE instructions to speed up the global sequence alignment \citep{ksw2}. 
Importantly, we report all our alignments in standard BEDPE format, together with corresponding edit strings in CIGAR format~\citep{samtools} and various other useful metrics similar to  WGAC such as Kimura two parameter genetic distance \citep{Kimura72} and Jukes-Cantor distance \citep{Jukes69}.

In our experiments, we used $k=12$ for the seed SD stage and $k=11$ for chaining step (note that this parameter is configurable by user). While lower values of $k$ may improve the sensitivity, we found that any such improvement is rather negligible and not worth the increase in the running time. On the other hand, higher values of $k$ improve the running time while lowering the sensitivity.


\section{Results}
We evaluated SEDEF using the human reference genome (UCSC hg19) and mouse reference genome (UCSC mm8), and compared its calls to WGAC calls \footnote{Note that as mentioned in the Introduction it is not possible to run WGAC without Sun Solaris operating system, therefore we were not able to benchmark it ourselves. WGAC calls were obtained from \url{http://humanparalogy.gs.washington.edu} and \url{http://mouseparalogy.gs.washington.edu}}. WGAC calls are the current gold (and only) standard of SDs in both human and mouse genomes, and are used as segmental duplication annotations by UCSC Genome Browser.

In case of human genome, the entire process took around 10 CPU hours with the peak RAM usage of 7 GB in single-CPU mode. SEDEF is also highly parallelizable, and it took only 14 minutes for the whole process to finalize on 80 CPU cores. This is a significant improvement over WGAC, which takes several weeks to complete (private communication). Similar running times were observed in mouse genome, despite the fact that mouse genome contains significantly more repeats than human genome and thus necessitates longer running times \citep{she2008mouse}. Run times on a single CPU and 80 CPU cores when ran in parallel via GNU Parallel \citep{Tange2011a} are given in Table~\ref{tab:runtime}.

\begin{table}[htb!]
  \centering
\caption{Running time performance of SEDEF in single-core mode and multi-core mode on 80 Intel Xeon E7-4860 v2 cores at 2.60 GHz.}
\begin{tabular}{lrrr}
   & \multicolumn{3}{c}{\bf Human (hg19)} \\ \hline
  & \textbf{Total } & Seeding and Extending & Chaining and Aligning \\ \hline
1 core       
& \textbf{10h 30m}  & 7h 33m     & 2h 57m\\
80 cores     
    & \textbf{0h 14m}   & 0h 10m      & 0h 04m \\
\hline
  & \multicolumn{3}{c}{\bf Mouse (mm8)} \\ \hline
& \textbf{Total } & Seeding and Extending & Chaining and Aligning\\ \hline
1 core       
    & \textbf{13h 07m}  & 7h 53m     & 5h 14m \\
80 cores     
    & \textbf{0h 30m}   & 0h 10m      & 0h 20m  \\ 
\hline
\label{tab:runtime}
\end{tabular}
\end{table}

SEDEF initially detected around 2,250,000 seed SD regions in human genome. After the chaining process, the final number of SDs was reduced to $\approx$186,400. Finally, after filtering out the common repeats and other spurious hits, we report 67,882 final SD pairs that cover 219 Mbp of the human genome. 
This is a significant increase over WGAC data, which reports 24,477 SD pairs that cover 159 Mbp of the genome. Of this 60 Mbp increase in the duplication content, 30 Mbp belongs to regions in the genome without common repeats. Figure~\ref{fig:err} shows the genome coverage, together with size and error distribution of SDs found by SEDEF and WGAC. The majority of SEDEF SDs have cumulative error $\delta$ (with affine gap penalty) around 15\%. 
As for the mouse genome, SEDEF found 352,991 final SDs which cover 259 Mbp of the genome, as compared to 140 Mbp covered by 117,213 WGAC SDs. Of the additional 120 Mbp found by SEDEF, 45 Mbp belongs to non common repeat regions.
       
\subsection{Filter and alignment accuracy}

\subsubsection{Simulations}
\begin{figure*}
\centering
\begin{minipage}[t]{0.45\linewidth}
\includegraphics[scale=0.45]{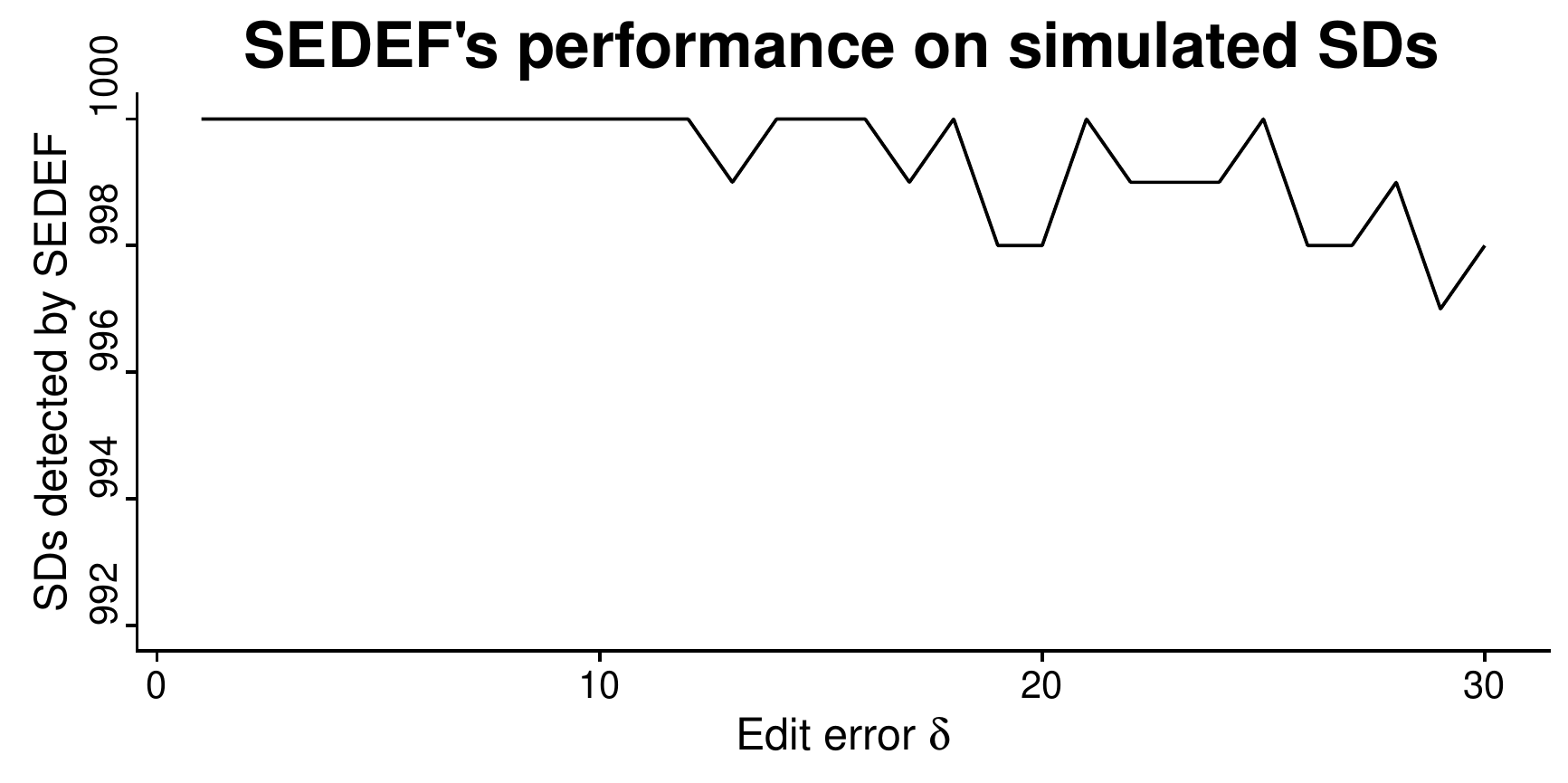}
\end{minipage}
\begin{minipage}[t]{0.49\linewidth}
\hspace{1em}\includegraphics[scale=0.19]{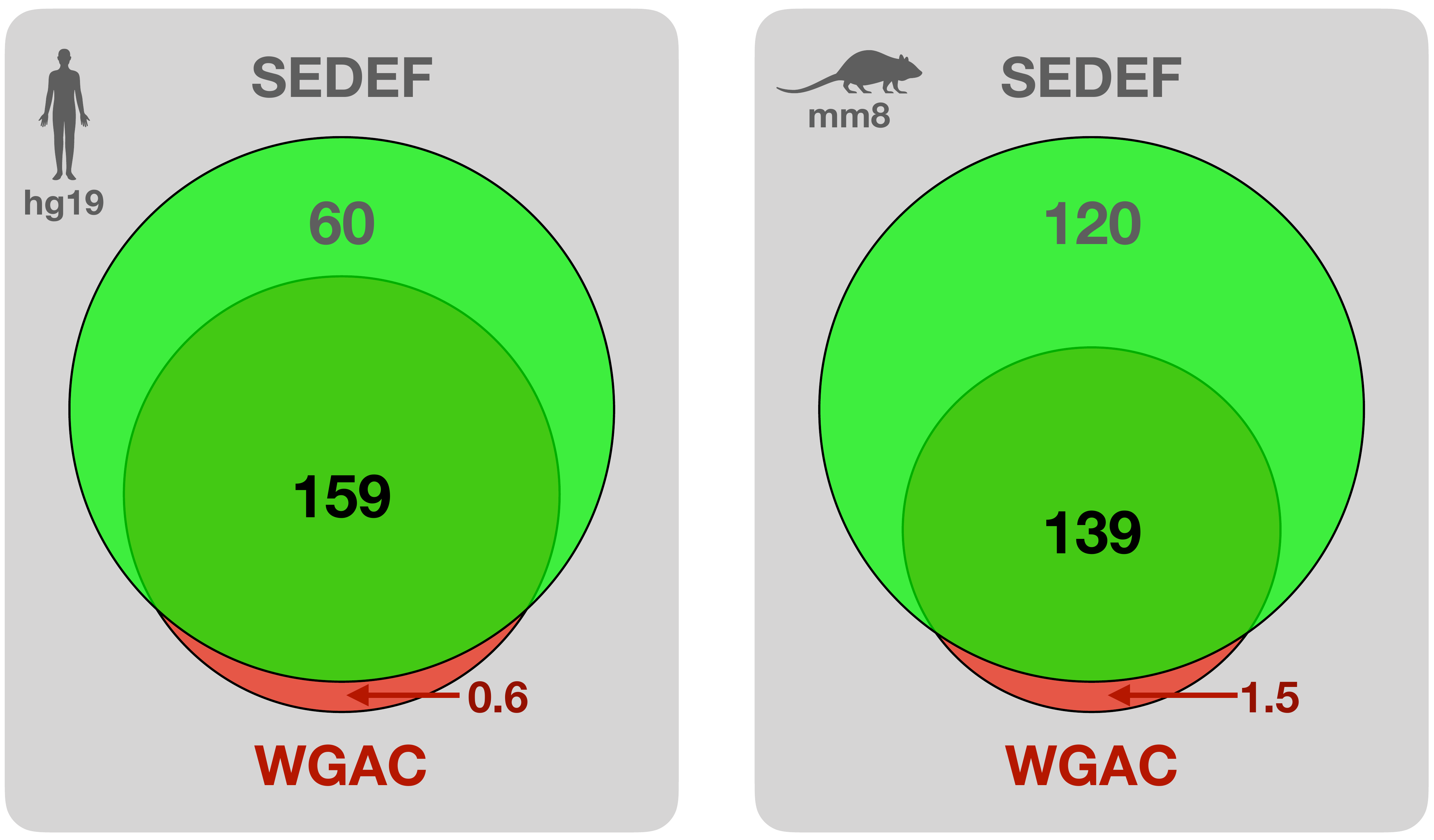}
\end{minipage}
\caption{\textbf{(Left)} Performance of SEDEF's algorithm on simulated SDs. $x$-axis is the total simulated SD error rate $\delta$, while $y$ axis is the number of correctly detected SDs (total 1,000 for each $\delta$). Since SEDEF successfully detects more than 995 simulated SDs for any $\delta$, the plot area is cropped.
\textbf{(Right)} Venn diagram depicts the SD coverage of the human and mouse genome (in Mbp) as calculated by SEDEF and WGAC. Intersected region stands for the bases covered by both SEDEF and WGAC.
} 
\label{fig:err}
\end{figure*}

We also evaluated the accuracy of the seeding and chaining process based on total error rate $\delta$. For this purpose, we generated 1,000 random sequences of sizes 1--100Kbp for each $\delta \in \{0.01,0.02,\dots,0.30\}$ (i.e. up to 30\%), and for each such sequence generated a random segmental duplication according to the SD criteria defined above (where $\delta_M$ and $\delta_G$ are randomly chosen such that they are both less than $\min\{0.15,\delta\}$). All sequences and mutations were randomly generated with uniform distribution. These two sequences (original one and the randomly mutated one) were fed to SEDEF, and then we checked whether SEDEF finds a match between these two sequences, and whether this match covers the original SDs (a match covers SD if more than 95\% of the SD bases are included in the match).
As shown in Figure~\ref{fig:err}, SEDEF's overall sensitivity is 99.94\%, and the sensitivity drops slowly as $\delta$ increases. However, even for $\delta = 0.30$, sensitivity remains above 99\%.

We performed a similar experiment on chromosome 1, where we randomly fetched 10,000 sequences (uniform distribution) of various lengths and introduced random mutations to simulate a SD event. In this experiment, SEDEF had only a 0.15\% false negative rate (i.e. undetected SDs), where all missed duplications  were very small SDs of lengths $\approx$1,000.

\subsubsection{WGAC coverage}
It is worth mentioning that SEDEF-detected human SDs completely cover $\approx$98\% of the previously reported SD intervals ($\approx$99.6\% in basepairs) by WGAC. SEDEF entirely misses less than 0.3\% (70) of SD intervals reported in WGAC results, and for the 1.4\% of WGAC SDs, SEDEF reports partial overlap (i.e. less than 80\% reciprocal overlap). All together, SEDEF misses about 0.6 Mbp out of 159 Mbp as reported by WGAC ($\approx$0.4\%), where 0.5 Mbp contained short common repeats.
We note that several WGAC SDs are in fact common repeats, and that several WGAC alignments contain long gaps. This is likely due to the dependency of WGAC on common repeat annotations, which may not be comprehensive. Additionally, WGAC employs several heuristics to reinsert common repeats to fuguized putative duplications that might ``glue'' very short non common repeat segments into larger segments with high repeat content that show similar alignment properties to a SD. 
This effect is much more present in mouse genome, where SEDEF misses 16,471 WGAC SDs (14.1\%), and partially covers 2.2\% of such SDs. However, in terms of basepairs SEDEF only misses 1.5Mbp (0.1 Mbp non common repeat elements). After extra validation, we found that most of missed WGAC SDs ($\approx$14,470) are in fact common repeats incorrectly reported as SDs; thus SEDEF misses only 1.7\% of the correct WGAC SD calls. 

\subsection{Comparison to other methods}

We also evaluated the SD discovery accuracy of whole-genome aligners Minimap2 \citep{Li2018} and MUMmer/nucmer \citep{marccais2018mummer4} on the human genome assembly (UCSC hg19). These tools do not support SD detection out of the box; however, a self assembly-to-assembly comparison can be performed in order to identify the repetitive regions in the genome. These regions can be refined into SDs after applying further processing with SDDetector \citep{dallery2017gapless} and filtering out candidate SDs which consist solely of common short repeats.
Compared to these tools, SEDEF is an integrated pipeline for identifying SDs from scratch on a given assembly. Note that other similar tools, such as DupMasker~\citep{jiang2008dupmasker}, are developed to annotate segmental duplications and require already existing SD database from similar genomes to be able to mark SDs in a given genome.

We ran these tools on 20 CPU cores using the GNU Parallel~\citep{Tange2011a} by aligning all pairs of chromosomes in hg19. Minimap2-based analysis identified only 29\% 
of the SD intervals reported by WGAC, which spanned 33\% of the duplicated basepairs (53 Mbp out of 159 Mbp). MUMmer/nucmer approached better SD coverage performance, which identified 98.8\% of WGAC regions 
that spanned 89\% of duplicated basepairs (143 Mbp out of 159 Mbp), but the analysis was much slower and completed in 20 hours in the same compute setting. Minimap2 required 1.5 hours of run times using 20 CPU cores (in comparison, SEDEF takes only 36 minutes on 20 cores). 

Overall, Minimap2-based analysis misses a significant amount of duplications in a self-comparison task when using the recommended parameters of intra-species assembly-to-assembly comparison. Meanwhile MUMmer/nucmer-based approach covers the SD regions more consistently with those reported by WGAC; however it still misses many WGAC calls which are found by SEDEF. Finally, SEDEF is able to find more calls compared to the other tools in much shorter amount of time, as shown in Table~\ref{tab:compare}.

\begin{table}[htb!]
\centering
\caption{SD coverage of the human genome (hg19) as reported by different tools. }
\begin{tabular}{lrrrr}
\textbf{Tool} & \textbf{Covers} & \textbf{Misses} & \textbf{Extra} & \textbf{Time (h:m)} \\
\hline
WGAC (gold standard) & 159.5 & 0.0 & 0.0 & weeks \\
SEDEF         & 218.8 &   0.6 & 60.0 & 0:36 \\
Minimap2      &  53.3 & 107.3 &  1.1 & 1:30 \\
MUMmer/nucmer & 142.6 &  30.8 & 13.9 & $\geq$20:00 \\
SDDetector    &  30.1 & 130.8 &  1.5 & $\geq$1:00* \\
\hline
\end{tabular}
\begin{minipage}{\textwidth}
\footnotesize
Misses and Extra are calculated with respect to the WGAC SD calls, which are currently the gold standard of SD calls. Note that we  have filtered out all calls where at least one mate is composed solely of common short repeats (Minimap2, MUMmer/nucmer and SDDetector) as we did on SEDEF. All running times were adjusted for 20 CPU cores (all tools which support parallelization were run on 20 cores). \\
$^*$Adjusted running time for 20 cores; in reality, SDDetector spends
$\geq 8$ hours in the single threaded pre-processing stage. Furthermore,
the reported running time only includes post-processing and does
not include initial BLAST alignment calcuations.
\end{minipage}
\label{tab:compare}
\end{table}


\section{Conclusion}
 Segmental duplications are among the most important forms of genomic rearrangements that drive genome evolution. However, their accurate identification is lacking due to the unavailability of necessary computational tools. In this manuscript we presented SEDEF to help fill this gap in methodology.
 
 In future work, we aim to characterize the effect of various edit distance embeddings and  techniques such as gapped $q$-grams \citep{Burkhardt2002,BarYossef2004}. While many of these techniques have been previously implemented \citep{Hanada2011}, our initial experiments did not show that any such embeddings or techniques are beneficial for strings with large edit distances. 
 
 SEDEF is designed as a fast, accurate, and user friendly tool to discover duplicated segments in genome assemblies. Therefore it aims to help researchers easily identify duplicated segments in genomes from several organisms, enabling them to extend their ability to perform comparative genomic studies in complex regions of the genome. We aim to extend it with an A-Bruijn graph based analysis~\citep{Jiang2007} to provide a full view of the evolution of segmental duplications.
 Armed with the extensions as we mention above, we will then use SEDEF to fully analyze reference genome assemblies from various genomes to both evaluate the assembly accuracy, and to better understand the role of segmental duplications in organism evolution.
 
\section*{Acknowledgements:} 

We thank  Evan E. Eichler for early discussions on formulating the problem, and Ashwin Narayan for helpful suggestions. \emph{Funding:} This work is supported in part by NSERC Discovery Grant to F.H., EMBO Installation Grant (IG-2521) to C.A. and NIH grant GM108348 to B.B. 

\subsubsection*{Conflict of interest:} None declared.

\end{document}